\documentclass{article}
\usepackage{fancyhdr}
\usepackage{graphicx}
\usepackage{amsmath, amsthm, amssymb}
\usepackage{gensymb}
\date{}
\begin{document}
\begin{center}
\textbf{\LARGE Analysis of Quantum Particle Automata for Solving the Density Classification Problem}\\
\vspace{0.5cm}
\textbf{Tina Yu \footnote{Tina Yu, Del Mar, California 92014, USA, e-mail: gwoingyu@gmail.com} and Radel Ben-Av 
\footnote{Radel Ben-Av, Software Engineering Department, Azrieli College of Engineering Jerusalem, POB 3566, Jerusalem, 91035, Israel}}
\end{center}

\begin{abstract}
\noindent
To advance our understanding of Quantum Cellular Automata in problem solving through parallel and distributed computing, this research quantized the density classification problem and adopted the Quantum Particle Automata (QPA) \cite{meyer} to solve the quantized problem.  In order to solve this problem, the QPA needed a unitary operator to carry out the QPA evolution and a boundary partition to make the classification decisions. We designed a Genetic Algorithm (GA) \cite{holland} to search for the unitary operators and the boundary partitions to classify the density of binary inputs with length 5. The GA was able to find more than one unitary operator that can transform the QPA in ways such that when the particle was measured, it was more likely to collapse to the basis states that were on the correct side of the boundary partition for the QPA to decide if the binary input had majority density 0 or majority density 1. We analyzed these solutions and found that the QPA evolution dynamic was driven by a particular parameter $\theta$ of the unitary operator: a small $\theta$ gave the particle small mass hence fast evolution while large $\theta$ had the opposite effect.  While these results are encouraging, scaling these solutions for binary inputs of arbitrary length of $n$ requires additional analysis, which we will investigate in our future work. 
\end{abstract}
\section{Introduction}
The density classification problem is an important test case to measure the computational power of Cellular Automata (CA)\cite{packard}\cite{mch}. In the CA computational model, there is a grid of cells where the states of the cells are updated synchronously according to a local rule. The density classification problem was proposed to study one-dimensional classical CA, where each cell contains a state of 0 or 1, and the task is to identify if the majority of the cells is 0 or 1. A solution to this problem is a local rule that converges the CA to a fixed point of all 1s if its initial configuration contains more 1s than 0s and to all 0s if its initial configuration contains more 0s than 1s. Such convergence normally takes place within \textit{M} time steps, where in general \textit{M} depends on the length of the lattice \textit{L} (which is assumed to have periodic boundary conditions, resulting in a circular grid). Over the years, various CA local rules have been proposed \cite{gkl}\cite{mch}\cite{abk}\cite{cst}  to demonstrate that classical CA can achieve global synchronization through parallel local interactions. These results indicate that the CA computational model is well suited for distributed and parallel computing. 

Meanwhile, extensions of classical CA to Quantum CA (QCA) for distributed and parallel computing have also been investigated \cite{grossing_zeilinger}\cite{meyer}. Unlike classical information of binary 0 \emph{or} 1, quantum information resides in superpositions of 0 \emph{and} 1 simultaneously. Computation takes place on each of the superpositions following a distinct path, called "quantum parallelism" \cite{deutsch}. Therefore, there is a natural mapping between the parallel computation on \textit{quantum superpositions} and  the parallel processing of classical \textit{CA cells}. In terms of distributed computing, Feynman and others  \cite{feynman}\cite{deutsch} have shown that quantum formalism of closed, locally interacting microscopic systems are able to perform universal computation. 

To advance our understanding of QCA in problem solving through parallel and distributed computing, this research quantized the density classification problem and adopted the Quantum Particle Automata (QPA) devised by Meyer \cite{meyer} to solve this problem.  

The QPA is a one-dimensional CA that consists of a \emph{single particle}. Similar to the classical CA, the state of each cell in the QPA at a given time step depends on the states of the cells in some local neighborhood at the previous time step. However, the evolution of the QPA is quantum mechanical. More precisely, unlike classical CA where the state of each cell $x_i$ is a binary value of 0 or 1, the state of a QPA cell is a (complex) probability amplitude $c_i$ for the particle being in state $|x_i\rangle$ when it is measured (being in state $|x_i\rangle$ means being in position $x_i$).  The state of the QPA $|\psi\rangle$ is a linear combination of all $L$ possible states, where $L$ is the length of the lattice: 
\begin{align*}
|\psi\rangle=c_0|x_0\rangle+c_1|x_1\rangle+\cdots+c_{L-1}|x_{L-1}\rangle
\end{align*}
Moreover, the QPA local rule that is used to update the probability amplitude in each cell is a unitary operator.  Since the transition of the QPA cells is unitary, the total probability, i.e. the sum of the norm squared of the probability amplitude at each cell, is preserved.

To solve the quantized density classification problem,  the QPA needs a unitary operator to carry out the QPA evolution and a boundary partition to make the classification decisions. This research applied Genetic Algorithms (GA) \cite{holland} to discover the unitary operators and the boundary partitions for the QPA to classify the density of binary inputs with length 5. GA is a population-based search algorithm that has been widely used in optimization and machine learning \cite{goldberg}. Through the simultaneous search of a population of candidate solutions, the GA was able to discover more than one unitary operator that can transform the QPA in ways such that when the particle was measured, it was more likely to collapse to the basis states on the correct side of the boundary partition for the QPA to decide if the binary input had density 0 or density 1. We analyzed these solutions and found that the QPA evolution dynamic was driven by a particular parameter $\theta$ of the unitary operator: a small $\theta$ gave the particle small mass hence fast evolution while large $\theta$ had the opposite effect.  While these results are encouraging, scaling these solutions for binary inputs of arbitrary length of $n$ requires additional analysis, which we will investigate in our future work.  

In addition to being the first to investigate QCA in solving the density classification problem, this research also made the following contributions:
\begin{itemize}
\item It devised a quantum version of the density classification problem that can be used to represent the problem for binary inputs of any length $n$.
\item It demonstrated that for binary inputs of length 5, there are multiple solutions to the quantized density classification problem and the GA methodology can find many of these solutions.
\item It analyzed these solutions and showed that  the QPA evolution dynamic is driven by a particular parameter $\theta$ of the unitary operator: a small $\theta$ gives the particle small mass hence fast evolution while large $\theta$ has the opposite effect.
\end{itemize}

The rest of the paper is organized as follows. Section \ref{qca} explains classical CA, quantum CA and the QPA that we used to conduct our research.  In Section \ref{density}, we first review the local rules proposed to solve the classical density classification problem and then quantize the problem for our QPA to solve this problem for binary inputs of length 5. Section \ref{simulation} describes the GA system we designed to discover the unitary operator and the boundary partition solutions. The results are then presented in Section \ref{result}. In Section \ref{analysis}, we analyze the two most extreme case unitary operator solutions and discuss scaling these solutions to binary inputs of arbitrary length $n$. Finally, Section \ref{conclusion} concludes this paper and outlines our future works.

\section{Quantum Cellular Automata} \label{qca}
In classical CA, there is a finite set of states $\Sigma$ and an infinite or finite lattice of $L$ cells, each of which is in one of the states in $\Sigma$. At each discrete time step $t$, the state of the lattice evolves according to some local rule, which transforms the state of each cell based on the state of some neighborhood cells at time step $t-1$. For example, \cite{gkl} and \cite{mch} employed a classical CA with two possible states ( $\Sigma=\{0,1\}$) and $L=149$ to solve the density classification problem. We will discuss the proposed classical local rules in Section \ref{density}.

CA updating is discrete in time and space. It is also space and time homogeneous, which means that at each time step the \emph{same local rule} is applied to update \emph{all cells} synchronously.  
When Gr$\ddot{o}$ssing and Zeilinger \cite{grossing_zeilinger} formulated the first QCA, they found that except for the trivial case, strictly local, unitary evolution of the whole QCA is impossible. In fact, Meyer \cite{meyer} later proved that ``there is no homogeneous one-dimensional QCA that is nontrivial, local and scalar." Gr$\ddot{o}$ssing and Zellinger therefore relaxed the unitary constraint in their QCA by allowing approximate unitarity during states updating.  After the updating at each time step, the states of the cells are normalized to make their QCA evolution unitary. This extra step also caused non-local interaction, which makes their QCA evolution non-local and the quantum unitiary non-linear.

An alternative QCA formulation approach is to relax the homogeneity constraint by partitioning CA \cite{watrous}\cite{meyer}. The main idea of partitioned CA \cite{toffoli_margolus} is that the set of cells are partitioned in some periodic way where every cell belongs to exactly one block partition. At different time steps, the local rule acts on a different block partition of the lattice. Such a QCA is neither time homogeneous nor space homogeneous anymore, but periodic in time and space. However, the quantum unitarity property can be maintained by using a unitary operator to update each block partition. Since unitary operators preserve the norm squared of the probability amplitudes in each block, the evolution of the entire QCA is unitary.

\begin{figure}[!htp]
\centerline
{\includegraphics[width=1.5in,height=1.0in]{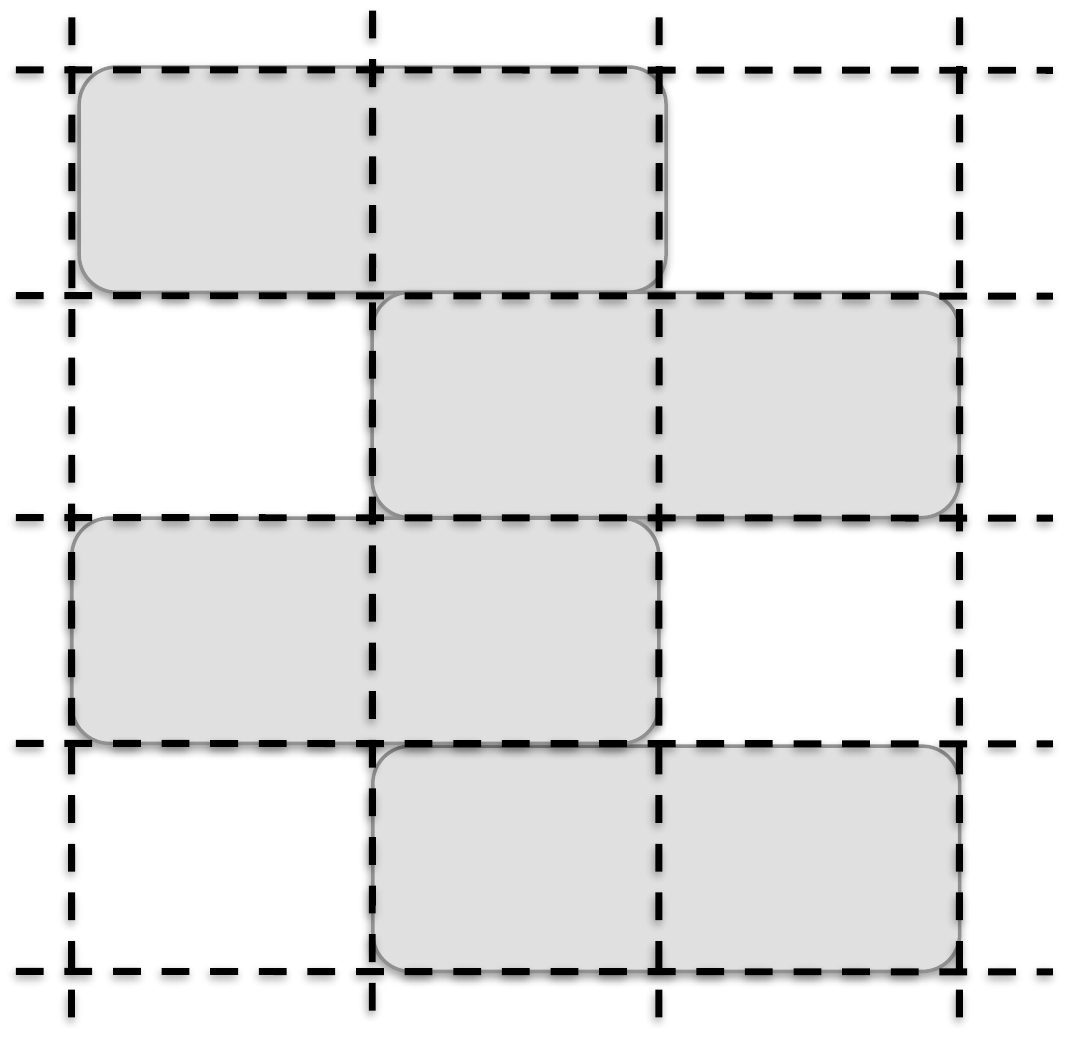}}
\centering
\caption{In QPA, each block partition contains 2 adjacent cells with the pairing changing at alternating time step. A unitary matrix is applied to each pair of the cells to update the states of the QPA.}
\label{partition}
\end{figure}

The QPA we used in this research is a partitioned QCA, where each block contains a pair of adjacent cells, with the pairing changing at alternating time step (see Figure \ref{partition}). A $2\times2$ unitary matrix is applied to each pair of the cells to update the states of the QPA. Hence, the QPA is  2-step translation invariant. 

The unitary operator used to update the QPA, $S \in U(2)$, is as follows:

\begin{equation}
S(\theta,\alpha,\beta,\gamma,\delta)=\begin{bmatrix} e^{i\alpha} sin \theta & e^{i\beta} cos \theta  \\ e^{i\gamma} cos \theta & e^{i\delta} sin \theta \end{bmatrix}
\label{unitary}
\end{equation}
where $(\alpha-\beta-\gamma+\delta)$ \% $2\pi \equiv \pi$; here \% is the modulo operator and $\equiv$ is the equal operator.

Updating the QPA for a single time step is achieved by a series of matrix multiplications on each pair of the cells:
\begin{equation}
\begin{bmatrix} \phi_{t+1}(x-1) \\ \phi_{t+1}(x) \end{bmatrix}=\begin{bmatrix} e^{i\alpha} sin \theta & e^{i\beta} cos \theta  \\ e^{i\gamma} cos \theta & e^{i\delta} sin \theta \end{bmatrix} \begin{bmatrix} \phi_{t}(x-1) \\ \phi_{t}(x) \end{bmatrix}
\label{transition}
\end{equation}
where $\phi_{t}(x)$ is the state of cell $x$ at time step $t$.

The cell-pairing $(x-1,x)$ in Equation \ref{transition} changes at alternating time step with the initial $x=t$\%$2$. Since the lattice is a circular grid under periodic boundary conditions, at \emph{even} time steps, the pairs $(x-1,x)$  are: $(L-1,0), (1,2), (3,4) \dots (L-3, L-2)$, while at \emph{odd} time steps, the pairs $(x-1,x)$ are: $(0,1), (2,3), (4,5) \dots (L-2, L-1)$, where $L$ is the length of the lattice.

At each time step $t$, each cell $x$ of the lattice contains a complex number, $\phi_t(x)$, which is the probability amplitude for the particle being in position $x$ when it is measured.  
Since $S$ is unitary, the total probability i.e. the sum of the norm squared of the amplitude of each cell, is always 1:

\begin{align*}
\sum_{x=0}^{L-1}|\phi_t(x)|^2=1
\end{align*}
where $L$ is the length of the lattice.

\section{The Density Classification Problem}\label{density}

Since the density classification problem was first formulated, various classical CA local rules have been proposed to solve the problem. The first one was by Gace, Kurdymov and Levin (GKL) \cite{gkl}, which consists of 2 rules: 
\begin{itemize}
\item if $\phi_{t}(x)\equiv0$, $\phi_{t+1}(x)=$ majority of $\phi_{t}(x),\phi_{t}(x+1),\phi_{t}(x+3)$; 
\item if $\phi_{t}(x)\equiv1$, $\phi_{t+1}(x)=$ majority of $\phi_{t}(x),\phi_{t}(x-1),\phi_{t}(x-3)$; 
\end{itemize}
The GKL rule gives classification accuracy of approximately 80\% when the test cases were randomly generated from all possible initial CA configurations under $L=149$. 

Later, Mitchell, Crutchfield and Hraber\cite{mch} designed a GA while Andre, Bennett and Koza \cite{abk} applied a Genetic Programming \cite{koza} system to discover CA local rules that gave improved performance. However, under the current problem formulation, Land and Belew \cite{land_belew} proved that for a one-dimensional grid of fixed size $L$ and fixed neighborhood size $r\ge1$, there is no perfect solution to correctly classify all possible inputs. Capcarrere, Sipper and Tomassini \cite{cst} therefore modified the problem slightly with a different final output specification: instead of a fixed-point configuration of either all 0s (when 0 is the majority in the initial configuration) or 1s (when 1 is the majority in the initial configuration), the final CA configuration can be either a background of alternative 0 and 1, with one or more blocks of at least two consecutive 0s (when 0 is the majority) or 1s (when 1 is the majority). With such modification, this problem can be solved perfectly using either rule 184 or rule 226 defined under Wolfram's classification \cite{wolfram}. 

To investigate the QPA's ability in solving this problem, we quantized the density classification problem in the following ways: 
\begin{itemize}
\item The QPA lattice consists of $L$ cells, where each cell $x$ represents its equivalent binary input. In other words, it is the \emph{index}, not the \emph{content}, of the cell that corresponds to the binary input whose density the QPA is classifying. The binary inputs can be in binary code, for example, binary input 00011 is represented by cell 3 while binary input 01000 is represented by cell 8. Alternatively, the binary inputs can be in Gray code \cite{gray}, for example, binary input 00011 is represented by cell 2 while binary input 01000 is represented by cell 7. To classify the density of binary inputs of length $n$, the QPA has $L=2^n$.
\item The state of cell $x$ at time step $t$ ($\phi_t(x)$) is a complex number, which is the probability amplitude of the particle being in position $x$ when it  is measured. For example, if $\phi_t(3) = 0.2+0.3i$, the probability of the particle being in position 3 is $|0.2+0.3i|^2=0.13$. Of course, $\sum_{x=0}^{L-1}|\phi_t(x)|^2=1$. 
\item Initially, only the cell which represents the binary input to be classified has probability amplitude 1, while the rest $L-1$ cells have probability amplitude 0. For example, to classify the binary input 1000 in binary code, the initial QPA has cell 8 with probability amplitude 1, while the rest of the cells have probability amplitude 0.
\item After the QPA is evolved for $M$ time steps using a unitary operator $S$, the particle is measured. We assume that within the $L$ bases Hilbert space, there is a partition between the basis states that classify the density of the binary inputs as 0 ($Z_{cells}$) and those that classify the density of the binary inputs as 1 ($O_{cells}$). Hence, if the particle collapses to one of the zero-state cells ($Z_{cells}$) more often ($>$ 50 \%) than it collapses to one of the one-state cells ($O_{cells}$), the QPA classifies the density of the binary input as 0. Otherwise, it classifies the density of the binary input as 1.  We will discuss how the partition is decided in the next section.
\end{itemize}

This is very different from the classical version of the density classification problem, where a CA is expected to converge to all 0s or all 1s. Since the QPA is a reversible CA, different initial configurations can never converge to identical states. Moreover, the QPA evolution is unitary, hence preserves the sum of the lattice states at each time step. This is not the case in classical CA, where the sum of the lattice states may vary from one time step to another.  

Under this quantization, the density classification problem has a general classical solution for binary inputs of any length $n$:
\begin{align*}
S=I=\begin{bmatrix} 1 & 0 \\ 0 & 1 \end{bmatrix} 
\end{align*}
\begin{align*}
 Z_{cells}=\{x: 0 \le x < n; majority(x_{binary})\equiv 0\} \\
 O_{cells}=\{x: 0 \le x < n; majority(x_{binary})\equiv 1\}
 \end{align*}
Since the identity operator (I) does not change the QPA initial configuration, by assigning all cells $x$ whose binary representation has majority 0 to $Z_{cells}$ and the others to $O_{cells}$, the particle always collapse to members of $Z_{cells}$ with certainty when the binary input has majority 0 and to members of $O_{cells}$ with certainty when the binary input has majority 1.

But classical solutions are not interesting. We would like to know if there is a quantum solution that can manipulate the complex probability amplitudes of the QPA to distinguish binary inputs with majority 0 from those with majority 1. The next section presents a GA that we designed to search for quantum solutions.

\section{Genetic Algorithms System Design}\label{simulation}
The GA we implemented searches for quantum solutions $S$ that solve the quantized density classification problem for binary inputs of length 5. 
As shown in Equation \ref{unitary}, $S$ has 5 parameters: $\theta, \alpha,\beta, \gamma, \delta$. Since $\alpha, \beta, \gamma, \delta$ are constrained, $(\alpha-\beta-\gamma+\delta)$ \% $2\pi\equiv\pi$, we only need to know 3 of these 4 parameters to define $S$. The number of the $S$ parameters that GA needed to optimize was therefore 4: $\theta, \alpha,\beta,\gamma$.

Initially, we specified the $Z_{cells}$ containing all odd cells of the lattice while the $O_{cells}$ containing all even cells of the lattice. Under this setup, the GA was not able to find a $S$ that can evolve the QPA to classify the density of all $2^5=32$ binary inputs correctly. We then made another attempt by specifying the left-half cells of the lattice as $Z_{cells}$ while the right-half cells of the lattice as $O_{cells}$ but the GA was still unable to find a correct $S$ solution. 
We therefore decided to let the GA to find the partition $Z$. Consequently, the total number of parameters that the GA optimized became 5: $\theta, \alpha,\beta, \gamma, Z$.

The partition $Z$ is represented as an integer value between 0 and $2^{32}-1$. To divide the 32 basis states into $Z_{cells}$ and $O_{cells}$, the 10-based integer value $Z$ is first converted into a 2-based binary string $Z_{binary}$.  For example,  $Z=4279011723_{10}$, $Z_{binary}=11111111000011001000100110001011_{2}$. Next, the indices to all 0 bits in the $Z_{binary}$ become the members of the $Z_{cells}$ while the indices to all 1 bits in the $Z_{binary}$ become the members of the $O_{cells}$. In the above example, 
$Z_{cells}$=$\{2,4,5,6,9,10,12,13,14,16,17,20,21,22,23\}$. When the particle is measured at time step $t$, if $\sum |\phi_t(x_i)|^2 > 0.5, \forall i \in Z_{cells}$, the particle is more likely to collapse to one of the zero-state cells than to one of the one-state cells, hence the binary input is classified with majority 0. Otherwise, the binary input is classified with majority 1.

In GA, the parameters undergoing optimization are called genes and they are packed into a linear chromosome, called individual. As the first attempt using GA to search for quantum solutions $S$, we only investigated binary inputs of length 5 in this study. Hence, the fitness ($f$) of an individual ($\theta, \alpha,\beta, \gamma, Z$) is its ability to correctly classify the density of $2^5=32$ binary inputs ($BI$). The higher the $f$ is, the better the $S$ and $Z$ are in solving the problem, i.e. this is a maximization problem.

\begin{equation}
\max f(\theta,\alpha,\beta,\gamma, Z) =\sum_{x=0}^{31} F(S(\theta,\alpha,\beta,\gamma), Z, BI_x)
\label{fitness}
\end{equation}
Here, $F(S(\theta,\alpha,\beta,\gamma), Z, BI_x$) is a function that returns 1 if $S$ classifies binary input $BI_x$ correctly based on the partition $Z$. Otherwise, F returns 0. An individual that correctly classifies the density of all 32 binary inputs would have $f=32$ and is a solution to the quantized density classification problem. 

Function $F$ operates in the following ways. First, partition $Z$ is used to generate $Z_{cells}$ and $O_{cells}$. Next, for each binary input $BI_x$, a QPA of lattice size $L=32$ is initialized with probability amplitude of 1 at cell $x$ and 0 at all other cells, i.e. $\phi_0(x)=1$ and $\forall_{y=0}^{31}, y \ne x, \phi_0(y)=0 $. At each time step $t$, $S(\theta,\alpha,\beta,\gamma)$ is applied to each cell block from time $t-1$ to produce new probability amplitudes for time step $t$. Once completed, the particle is measured. If $\sum |\phi_t(x_i)|^2 > 0.5, \forall i \in Z_{cells}$, the QPA classifies $BI_x$ with majority 0. Otherwise, it classifies $BI_x$ with majority 1. 
After that, the QPA classification is compared to the correct classification. If the two match, $F$ returns 1. Otherwise $F$ returns 0. The same process is repeated for all 32 binary inputs to obtain the fitness $f$. If $f$ is 32, indicating $S$ classifies all 32 binary inputs correctly, the QPA evolution stops.  Otherwise, the QPA continues to time step t+1. In other words, the number of time step $M$ of the QPA is not fixed but varies depending on $S$ and $Z$. The maximum allowed time step $M$ is 2,048 in this study\footnote{In physical implementation, when a particle is measured, it collapses to one of the possible states. Meanwhile, all superposition information is destroyed by the measurement operation. Since the objective of this research is to solve the quantized density classification problem by simulating the algorithm in a classical computer, the measurement of a particle does not destroy the amplitudes at each superpositions. Once a solution ($S, M, Z$) is obtained, the physical implementation of a particle only requires one measurement (at time step $M$) to solve the problem.}. 

In our implementation (see Appendix A), function F only evolved QPA once and the probability amplitudes in the 32 cells can be used to classify all 32 binary inputs. This is possible because the lattice is a grid with periodic boundary conditions. As a result, to classify each binary input, we only need to shift the grid one cell left and perform the same evaluation procedure to obtain the QPA classification. In other words, to classify binary input $BI_x$, cell number $(32-x)\%32$ is treated as cell 0 to test against $Z_{cells}$ for QPA classification decision.

The GA uses the following operators to optimize $S$ and $Z$:
\begin{itemize}
\item Uniform crossover: each individual is paired with a randomly selected different individual (partner) in the population to produce one offspring. The gene values of the offspring are decided in the following ways. A random number is generated for each of the 5 genes; if the random number is less than the crossover rate ($cR$), the gene is selected from the original individual. Otherwise, it is selected from the partner individual. The smaller the $cR$ is, the higher portion of the original individual is disrupted to produce its offspring.

\item Gaussian mutation: the gene values of an individual is mutated in the following ways.  A random number is generated for each of the 5 genes; if the random number is less than the mutation rate ($mR$), the gene is mutated by adding a random value from the Gaussian distribution under the standard deviation ($std$) specified for that gene. Otherwise, the gene value remains the same. The higher the $mR$ is, the higher portion of the individual is disrupted with new gene values.

\item Survival selection: if an individual is better than its offspring, i.e. with higher fitness $f$ according to Equation \ref{fitness}, the individual survives to the next generation. Otherwise, it is replaced by its offspring.
\end{itemize}

\begin{figure}[!htp]
\centerline
{\includegraphics[width=5in,height=4.5in]{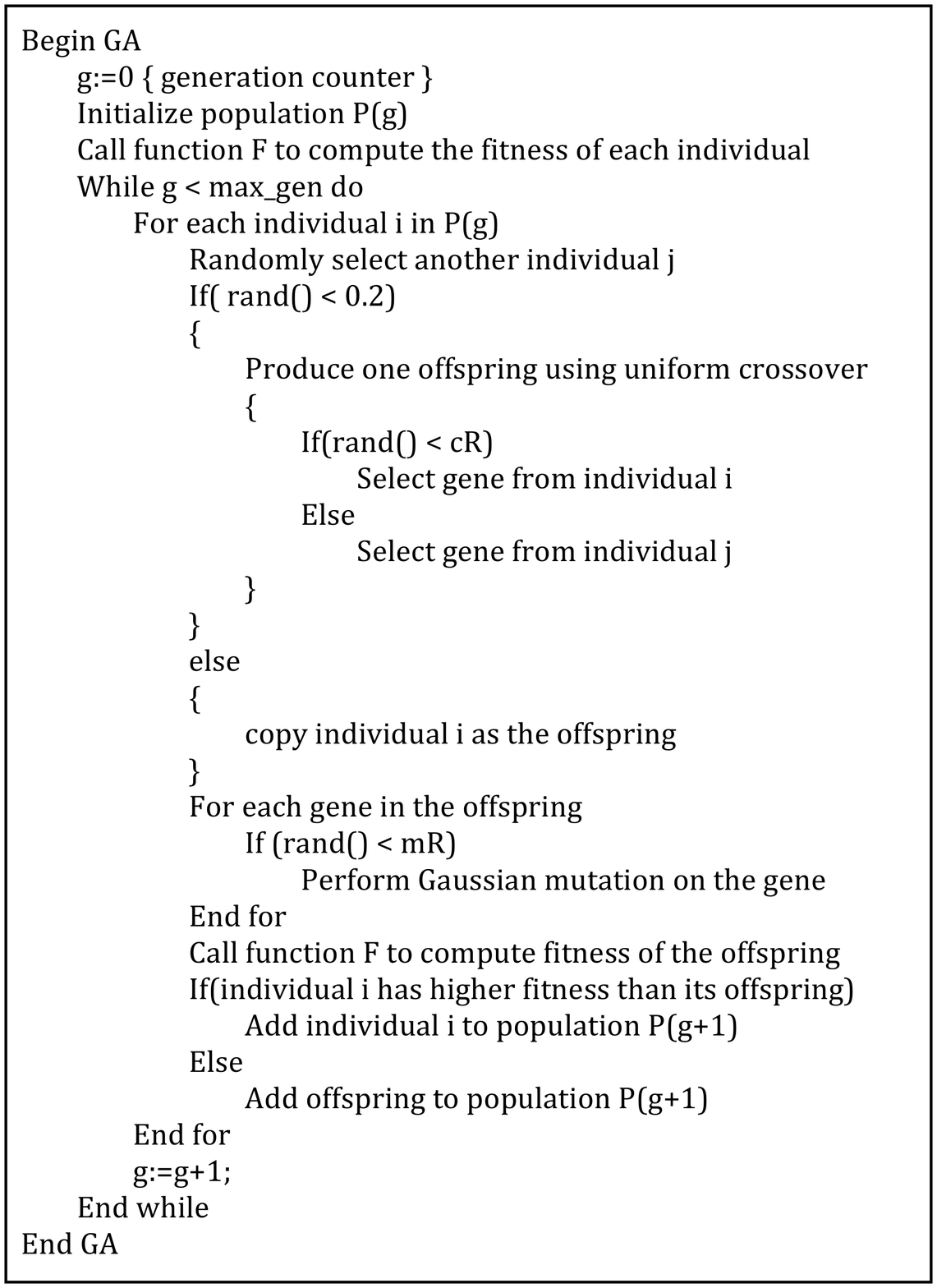}}
\centering
\vspace{-1.5cm}
\caption{The GA workflow to discover unitary operators S and partition Z for the QPA to solve the quantized density classification problem.}
\label{ga}
\end{figure}

Figure \ref{ga} gives the work flow of the GA optimization process. Initially, a population of individuals, each contains ($\theta, \alpha, \beta, \gamma, Z$), is randomly generated. After applying function F to evaluate its fitness, each individual in the population undergoes either uniform crossover or Gaussian mutation alone or both to produce one offspring. Under the current setup, $\sim80\%$ of the offspring are produced from mutation alone, $\sim20\%$ of the offspring are produced from both mutation and crossover. Only $\sim0.0002\%$ of the offspring are produced from crossover alone. Next, the offspring's fitness is evaluated. If the individual is better than its offspring, it survives to the next generation. Otherwise, the offspring is selected to the next generation. This process will create a new generation of population that contains individuals whose fitness are either better or as good as those in the previous generation. After repeating this process for the specified maximum number of generations, the best individual in the last population is selected as the final solution.

We experimented with different $cR$ and $mR$ values and found that imposing higher disruption on individuals to produce offspring worked better in optimizing $S$ and $Z$. This might be due to the fact that our GA uses a survival selection scheme which technically is performing hill-climbing in the parameters space. Hilling-climbing is a greedy search algorithm. When all individuals perform this greedy search, the population diversity is reduced quickly and the search suffers from premature convergence. By using high $mR$ and low $cR$, a large amount of diversity is introduced to the population to balance the ``greediness"  of the survival selection, hence leads to more effective search.

\begin{table}[!htp]
\centering
\caption{GA system parameter values used to run the optimization.}
\begin{tabular}{|c|c|c|c|} \hline
{\bf parameter}&{\bf value} & {\bf parameter}&{\bf value}\\ \hline\hline
pop\_size&200& max\_gen&  1,000 \\\hline
xover\_rate (cR)& 25\% & mutation\_rate (mR) & 90\% \\\hline
$\theta$ range & $0\sim\pi/2$ & $\theta$ std & 0.05 \\\hline
$\alpha,\beta,\gamma$ range & $-\pi\sim\pi$& $\alpha,\beta,\gamma$ std & 0.1\\\hline
$Z$ range & $0\sim2^{32}-1$& $Z$ std & 20 \\\hline
\end{tabular}
\label{para}
\end{table}

Table \ref{para} lists the GA parameter values used to run the optimization. Additionally, we list the value ranges of $\theta, \alpha, \beta, \gamma$ and $Z$. In function $F$, these values are validated prior to be used to simulate the QPA for density classification. If any of the values is out-of-range, a random value within the legal range is generated to replace the out-of-range value.

\section{Results}\label{result}

We made multiple GA runs using both binary and Gray code representations of the binary inputs. It appeared that Gray code was more suitable for this problem and was easier for the GA to discover unitary operator solutions. Among the 50 Gray code runs, 16 found a solution while only 9 of the 50 binary code runs found a solution. This might be because under Gray code representation, neighboring cells always represent two binary inputs that are different in only 1 digit. As a result, the unitary operator $S$ always updates the complex probability amplitudes of 2 binary inputs that are different in 1 digit. Such consistency made it easier for the unitary operator to classify the density of the given binary input correctly.

\begin{table}[!htp]
\centering
\begin{tabular}{c|cccc}
\hline
run&$\theta(\degree)$&$\alpha(\degree)$&$\beta(\degree)$&$\gamma(\degree)$\\
\hline\hline
1&11.050685&	-55.57717&	113.034549&	-140.809934\\
2&8.0667&	-24.173737&	44.174208&	-129.677467\\
3&10.880958&	-19.904022&	104.97193&	-176.437282\\
4&75.337518&	5.369795&	43.854368	&-49.357459\\
5&7.259019&	-27.299086&	-62.015774&	111.243832\\
6&13.700568&	76.91783&	111.454715&	134.389406\\
7&23.718054&	-86.090944&	127.961266&	106.217157\\
8&25.152128&	170.66293	&-49.58397&	-5.4686\\
9&57.076225&	-111.908294&	88.120218&	-106.898772\\
\hline
\end{tabular}
\caption{The 9 unitary operator solutions found from the binary code runs.
\label{binarypara}}
\end{table}
\begin{table}[!htp]
\centering
\begin{tabular}{c|cccc}
\hline
run&$\theta(\degree)$&$\alpha(\degree)$&$\beta(\degree)$&$\gamma(\degree)$\\
\hline\hline
1&	4.407264&	119.105462&	-56.65684&	155.808914\\
2&	22.357464	&117.90735&	-9.633698&	151.621098\\
3&	10.431628&	-129.653366&	35.876261&	-37.38258\\
4&	7.303212&	-36.827341&	42.439786&	13.288965	\\
5&	12.565333&	-133.764371&	-116.992579&	27.211073	\\
6&	6.568701&	167.726702&	-105.341636&	125.939087	\\
7&	27.507306	& -72.340869&	-127.871776&	-23.952314	\\
8&	20.083639	&93.743705&	-44.820769&	-109.822316	\\
9&	27.284116&	-29.442762&	-49.124343&	32.082641	\\
10&	10.472256&	-63.163587&	11.258074&	-110.796705	\\
11&  28.257343&	41.031094	&-28.261072&	-83.989587	\\
12&	36.973794&	105.346366&	168.109796&	110.596441	\\
13&	12.472202&	-24.919773&	1.682202&	-178.873208	\\
14&	29.280147&	-124.541844&	120.74784&	62.525769	\\
15&	24.339945&	148.928022&	87.470762	&-106.736046	\\
16&	37.054707&	-22.12085&	-48.986319&	14.976938	\\
\hline
\end{tabular}
\caption{The 16 unitary operator solutions found from the Gray code runs.
\label{graypara}}
\end{table}
Tables \ref{binarypara} and \ref{graypara} give these unitary operator solutions ($\theta, \alpha, \beta, \gamma$) found in the binary and Gray code runs while Tables \ref{binarystep} and \ref{graystep} present the partition $Z$ solutions and the number of time steps $M$. As shown, the unitary operators have a wide range of values and they evolved the QPA for a different number of time step $M$. They also used a different $Z$ to classify the density of the binary inputs. In Tables \ref{binarystep} and \ref{graystep}, the row ``majority" gives the correct density classification for binary input $x$; they are listed from left to right based on cell order ($0\leq x\le 31$).  Similarly, $Z_{binary}$ gives the binary string converted from the partition $Z$, which is also listed in cell order from 0 to 31. 

One similarity among the $Z_{binary}$ solutions is that the number of 0-bit is close to half of the total number of binary inputs. This makes sense as half of the 32 binary inputs has majority 0 and the other half has majority 1. By allocating half of the cells to  members of $Z_{cells}$ and the other half to members of $O_{cells}$, it is easier for the unitary operator to balance probability amplitudes among members of $Z_{cells}$ and $O_{cells}$ to classify both majority-0 binary inputs and majority-1 binary inputs correctly. 


\begin{table}[!htp]
\centering
\begin{tabular}{|c||c||c||c|}
\hline
x&\scalebox{0.71}{0-1-2-3-4-5-6-7-8-9-10-11-12-13-14-15-16-17-18-19-20-21-22-23-24-25-26-27-28-29-30-31}&0-bits&\\
\hline\hline
majority&0-0-0-0-0-0-0-1-0-0-0-1-0-1-1-1-0-0-0-1-0-1-1-1-0-1-1-1-1-1-1-1&16&\\
\hline
run& $Z_{binary}$&&$M$\\
\hline \hline
1&1-1-0-1-0-0-0-1-1-0-0-1-0-0-0-1-0-0-1-1-0-0-0-0-1-1-1-1-1-1-1-1&15&2048\\
2&0-0-0-1-0-0-0-0-1-0-0-1-1-1-1-1-0-1-1-1-1-1-0-0-0-0-0-1-1-1-0-1&16&1368\\
3&0-1-1-1-0-1-0-0-0-1-1-0-0-0-0-0-0-1-1-1-1-0-1-1-0-1-1-0-0-1-0-0&17&1880\\
4&1-0-1-0-1-1-1-0-0-1-1-1-1-1-0-0-0-1-1-0-0-0-0-0-0-1-1-0-1-1-0-0&16&2034\\
5&1-1-1-1-0-0-0-1-0-0-0-1-1-0-0-1-1-1-1-1-1-1-0-0-1-1-0-0-0-0-0-0&16&2048\\
6&1-1-0-1-1-0-0-1-0-0-0-0-1-1-0-1-0-0-1-1-0-0-0-1-1-0-0-0-1-1-1-1&16&1184\\
7&0-1-0-0-0-1-1-0-0-0-0-0-0-1-1-0-1-0-1-1-0-1-1-1-1-1-0-0-0-1-1-1&16&1594\\
8&1-1-0-1-0-0-0-1-0-0-0-0-0-0-1-1-1-1-0-0-0-1-1-0-1-0-0-1-1-1-1-1&16&1182\\
9&0-1-0-1-1-0-0-0-0-0-0-1-0-0-0-0-1-1-1-1-1-1-0-1-1-1-0-1-1-0-0-1&16&768\\
\hline
\end{tabular}
\caption{The $Z_{binary}$ and the time step $M$ solutions from the binary code runs.}
\label{binarystep}
\end{table}

\begin{table}[!htp]
\centering
\begin{tabular}{|c||c||c||c|}
\hline
x&\scalebox{0.71}{0-1-2-3-4-5-6-7-8-9-10-11-12-13-14-15-16-17-18-19-20-21-22-23-24-25-26-27-28-29-30-31}&0-bits&\\
\hline\hline
majority&0-0-0-0-0-1-0-0-0-1-1-1-0-1-0-0-0-1-1-1-1-1-1-1-0-1-1-1-0-1-0-0&16&\\
\hline
run& $Z_{binary}$&&$M$\\
\hline \hline
1&1-0-0-1-1-0-1-1-0-0-0-1-1-1-0-1-1-0-0-0-0-1-0-0-1-0-0-1-0-1-1-1&16&2048\\
2&0-1-0-1-1-1-0-1-1-1-1-1-1-1-0-0-0-1-1-0-1-0-0-0-1-1-0-0-0-0-0-0&16&2012\\
3&1-0-1-1-1-0-1-1-1-1-1-1-0-0-0-0-1-1-0-1-1-0-0-0-1-0-0-0-0-0-0-1&16&811\\
4&0-1-0-0-1-1-0-0-0-0-1-0-0-0-0-0-0-1-0-1-1-1-1-1-1-1-0-1-1-1-0-0&17&2048\\
5&0-1-0-1-1-1-0-0-1-0-1-1-1-1-0-0-0-1-0-1-1-1-1-0-1-1-0-0-0-0-0-0&16&624\\
6&0-0-0-1-0-1-1-1-0-0-0-1-0-0-0-0-0-0-0-1-1-1-1-1-0-1-1-1-1-0-1-0&17&2046\\
7&1-1-0-0-1-0-0-0-0-1-1-1-1-0-0-0-0-0-0-0-1-1-0-1-1-1-0-1-1-1-0-1&16&1904\\
8&0-1-0-0-1-0-1-1-1-1-1-0-1-1-0-1-0-1-0-1-1-1-1-1-0-0-0-0-0-0-0-0&16&1274\\
9&0-0-0-0-0-0-0-1-1-1-1-1-0-1-1-1-1-1-1-1-0-0-0-1-0-1-1-1-0-0-0-0&16&620\\
10&0-1-1-1-1-0-1-1-0-1-0-1-0-0-0-1-1-0-1-0-0-0-0-1-0-0-1-1-0-1-1-1&15&1978\\
11&0-0-1-1-0-0-0-1-0-0-0-1-1-1-1-1-0-0-1-1-0-1-1-1-0-0-0-1-0-1-1-0&16&1572\\
12&1-1-1-0-0-1-0-1-1-1-0-0-0-1-0-0-1-0-0-0-0-1-0-1-0-1-0-1-1-0-1-1&16&632\\
13&1-1-1-1-0-0-0-0-1-0-0-0-0-1-0-1-0-1-1-0-0-1-1-1-1-1-0-0-0-0-1-1&16&1750\\
14&0-1-1-1-0-0-0-1-0-1-1-1-1-0-1-0-0-0-0-0-0-0-0-1-0-1-1-1-0-1-1-1&16&800\\
15&1-1-1-0-0-0-0-0-0-0-0-0-0-1-1-1-0-0-1-1-0-0-1-1-1-1-1-1-0-1-0-1&16&2048\\
16&0-1-1-0-0-1-0-1-1-1-1-1-1-0-0-1-1-1-0-0-1-1-0-0-1-1-0-0-0-0-0-0&16&684\\
\hline
\end{tabular}
\caption{The $Z_{binary}$ and the time step $M$ solutions from the Gray code runs.}
\label{graystep}
\end{table}
Since all these unitary operators classify the density of the 32 binary inputs of length 5 correctly, it is natural to ask ``do they produce the same QPA evolution dynamics, in terms of the probability amplitudes propagation?" 
We will analyze these unitary operator solutions through QPA simulations and answer this question in the following section.

\section{Analysis and Discussion}\label{analysis}

We simulated QPA evolutions and found that their dynamics were different under different unitary operator solutions. In particular, some QPA propagated the probability amplitudes very fast while others transformed the states in a much slower pace. Further analysis shows that the major driving force of the propagation is the value of $\theta$: a small $\theta$ gives the particle small mass hence fast propagation while large $\theta$ has the opposite effect. When $\theta=0\degree$, $S$ simply interchanges the states of adjacent cells so the probability propagation speed is 1 in lattice units. By contrast, when $\theta=\pi/2$, $S$ is the identity with phase in diagonal, where the phase is unobservable, hence there is no flow. Meyer \cite{meyer} has made similar observations in his QPA simulations under one-parameter ($\theta$) unitary operators. Since we used a more general unitary operator with 5 parameters, the QPA evolution dynamics were more complicated, given that $\alpha, \beta, \gamma, \delta$ also influenced the probability amplitudes propagation\footnote{By parameterizing $S$ from a $U(2)$ to a $SU(2)$, the number of independent parameters can be reduced to 4.}. We examined the QPA dynamics that were evolved under unitary operators that handled binary code and that handled Gray code representations and found that they had similar patterns. Thereafter, we will use the QPA evolved by unitary operators that handle binary code to perform our analysis.


\begin{figure}[!htp]
\vspace{-2cm}
\begin{minipage}[t]{0.49\linewidth}
\centerline{\includegraphics[width=7.8cm,height=9cm]{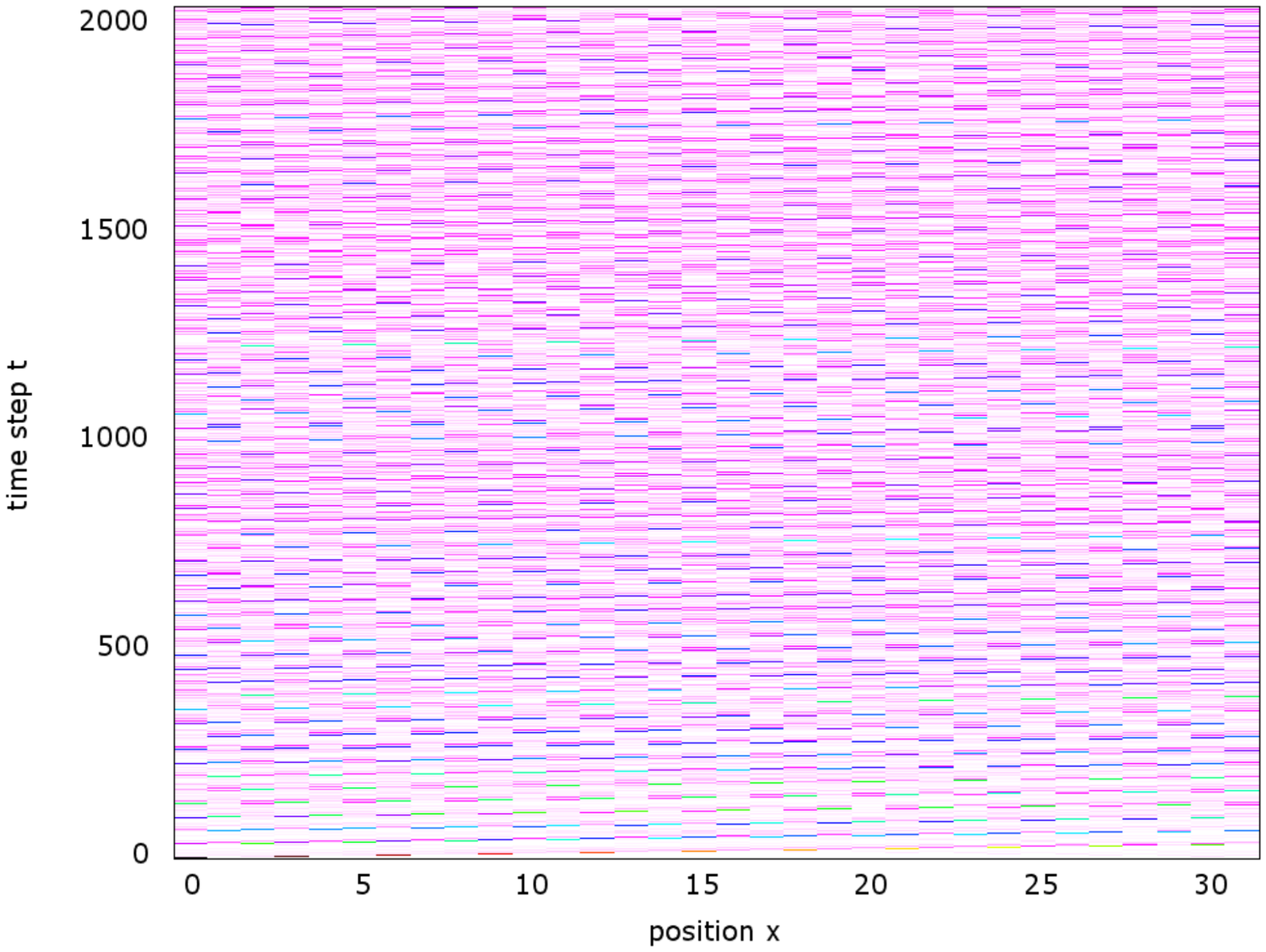}}
\vspace{-2cm}
\caption{QPA simulation with fast propagation under $S=\{\theta=7.26\degree,\alpha=-27.3\degree,\beta=-62\degree,\gamma=111.24\degree\}$; time step M=2048.
\label{fast}}
\end{minipage}
\hspace{-0.0cm}
\begin{minipage}[t]{0.49\linewidth}
\centerline{\includegraphics[width=7.8cm,height=9cm]{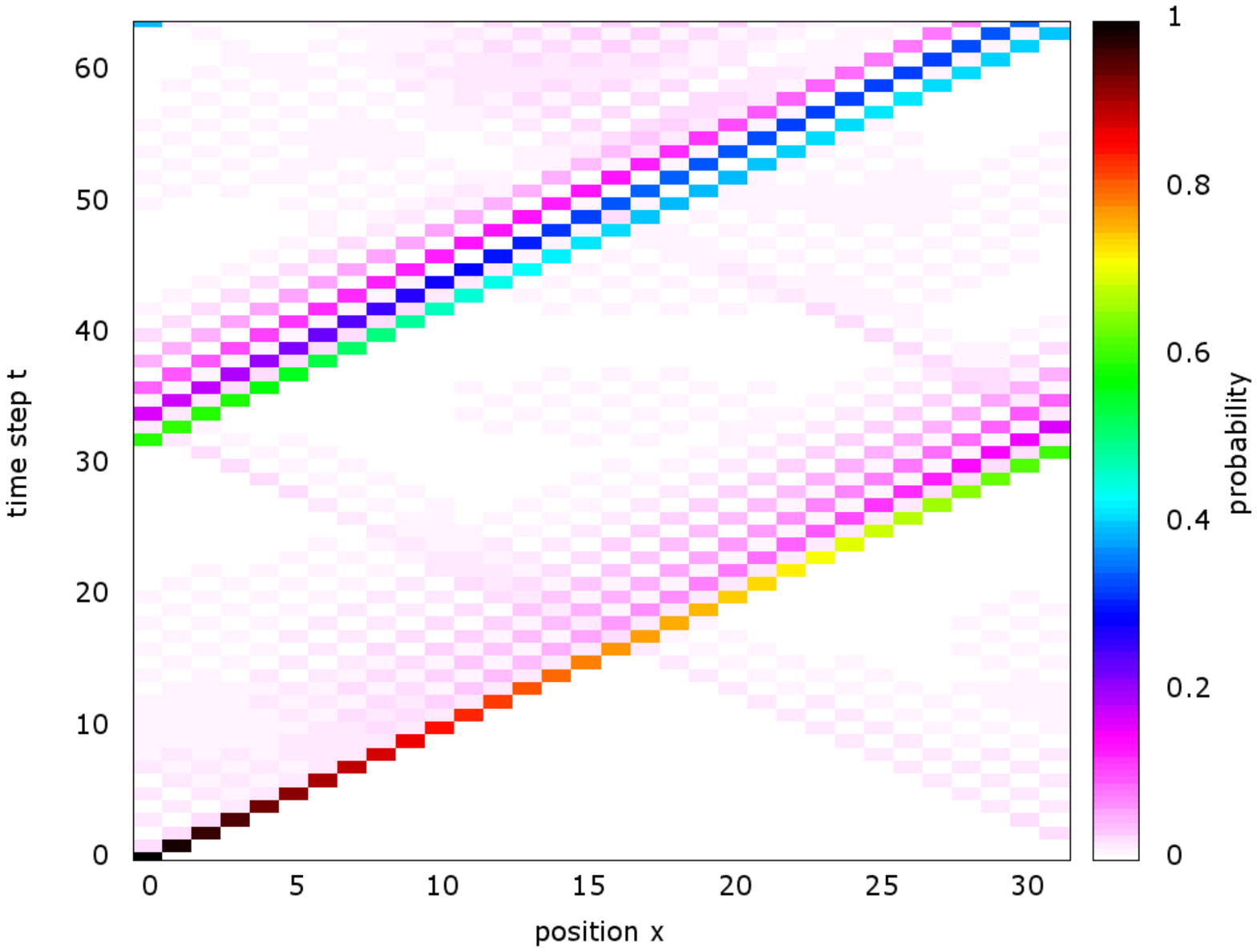}}
\vspace{-2cm}
\caption{QPA simulation under the same unitary operator as the plot on the left; first 64 time steps.
\label{fast64}}
\end{minipage}
\end{figure}

Figure \ref{fast} gives the QPA simulation that has the fastest propagation flow among all solutions.  The unitary operator is $S=\{\theta=7.26\degree,\alpha=-27.3\degree,\beta=-62\degree,\gamma=111.24\degree\}$. The value shown at each cell $x$ at time step $t$ is the norm squared of the probability amplitudes, i.e. $|\phi_{t}(x)|^2$. Figure \ref{fast64} shows the evolution during the first 64 time steps. Measuring the locations of the peaks of the probability distribution at successive time steps indicates that the propagation speed is a little bit less than 1 in lattice units.

 
\begin{figure}[!htp]
\vspace{-4.0cm}
\begin{minipage}[t]{0.49\linewidth}
\centerline{\includegraphics[width=7.5cm,height=9cm]{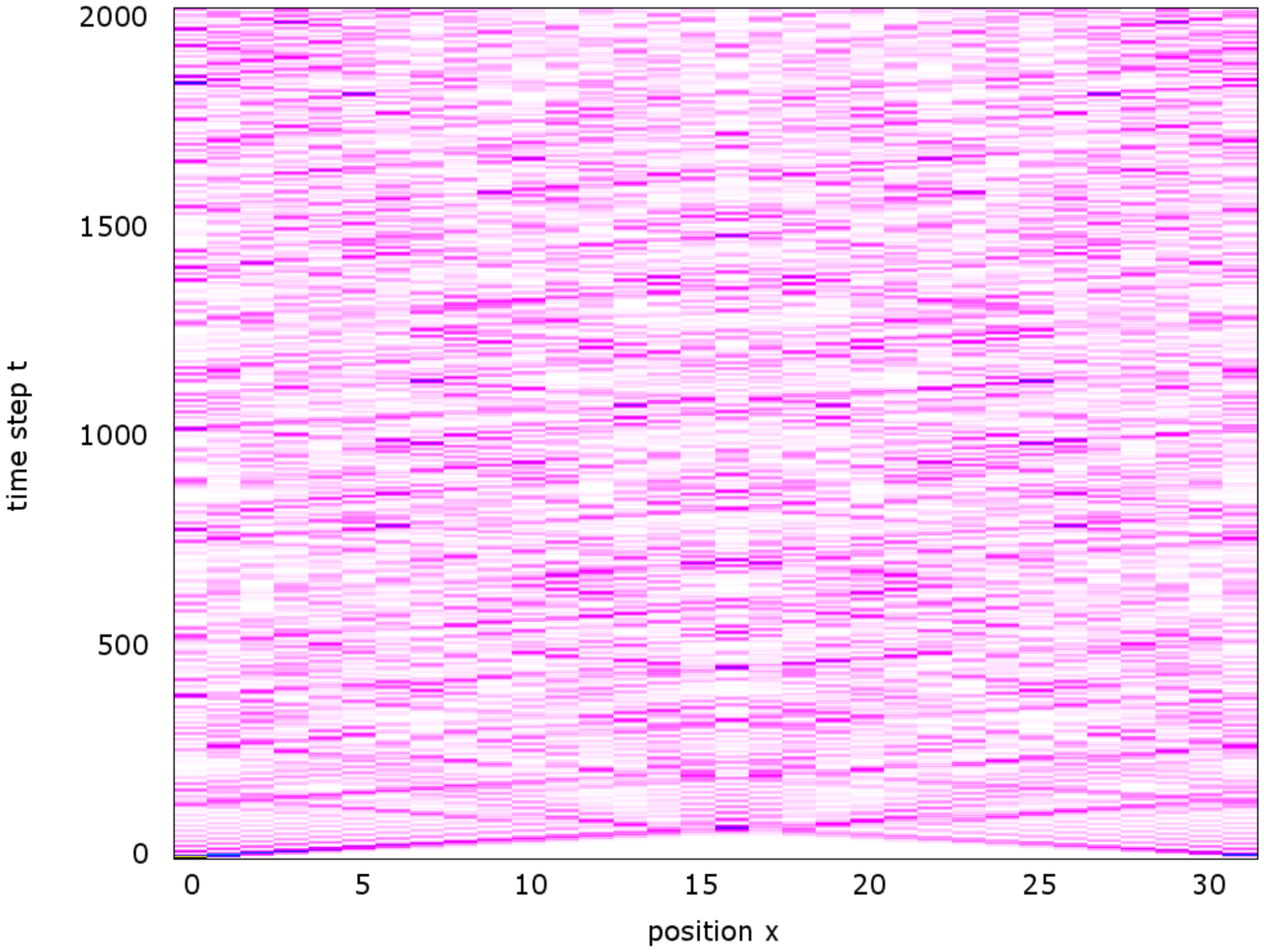}}
\vspace{-2cm}
\caption{QPA simulation with slow propagation under $S=\{\theta=75.3\degree,\alpha=-5.37\degree,\beta=43.84\degree,\gamma=-49.36\degree\}.$; time step M=2034.
\label{slow}}
\end{minipage}
\begin{minipage}[t]{0.49\linewidth}
\centerline{\includegraphics[width=7.5cm,height=9cm]{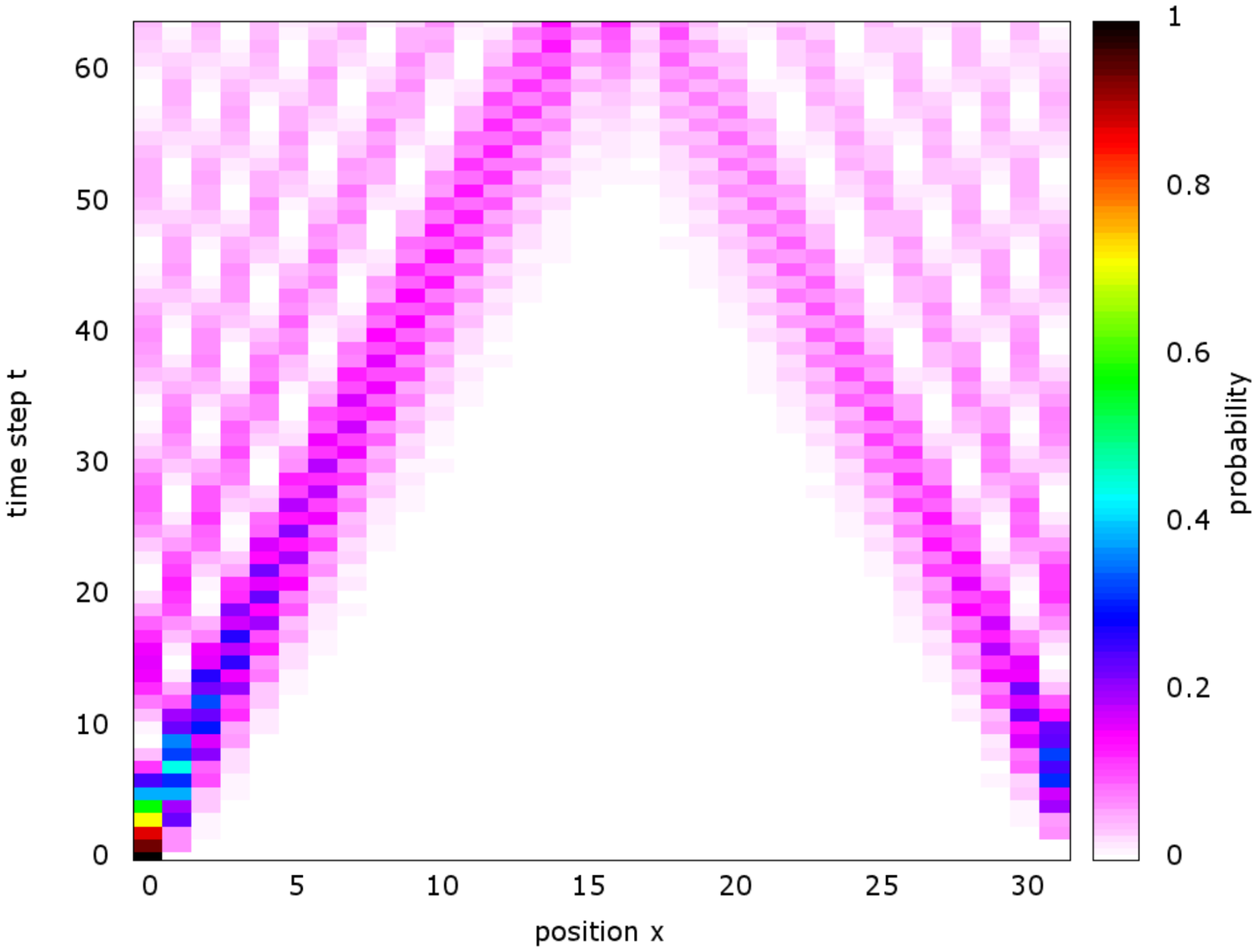}}
\vspace{-2cm}
\caption{QPA simulation under the same unitary operator as the plot on the left; first 64 time steps.
\label{slow64}}
\end{minipage}
\end{figure}

Similarly, Figure \ref{slow} gives the QPA simulation that has the slowest propagation flow among all solutions. The unitary operator is $S=\{\theta=75.3\degree,\alpha=-5.37\degree,\beta=43.84\degree,\gamma=-49.36\degree\}.$ Figure \ref{slow64} shows the probability propagation during the first 64 time steps. Measuring the locations of the peaks of the probability distribution at successive time steps indicates that the propagation speed is approximately  1/4 in lattice units.

When $\theta=\pi/2$ or $\theta=0\degree$, the QPA evolution dynamics are not interesting; the probability ($|\phi_{t}(x)|^2$) in each cell at the current time step is either an identical copy or a swap of its neighboring cell from the previous time step. 
However, when $0<\theta<\pi/2$, the probabilities in the QPA are propagated in a wide range of speeds, depending mostly on the $\theta$ of the unitary operator. Moreover, there are many unitary operators that can propagate the probabilities in ways such that when the particle is measured, it is more likely to collapse to the basis states that are on the correct side of the boundary partition $Z$ for the QPA to decide if the binary input has majority 0 or majority 1.

As mentioned in Section \ref{density}, the classical solution $S=I$ uses the $Z_{binary}$ that is the same as the \emph{majority binary string} to classify the density of the binary inputs. By contrast, quantum solutions use $Z_{binary}$ that are very different from the \emph{majority binary string} to classify the density of the binary inputs (see Table \ref{binarystep} and \ref{graystep}). In other words, within the 32 bases Hilbert space, there exists many partitions $Z$ between the basis states that classify the density of the binary inputs as 0 and those that classify the density of the binary inputs as 1. Many unitary operators can pair with these partitions to solve the density classification problem for binary inputs of length 5. However, in order for the unitary operator solutions to work for binary inputs of any length $n$, the paired partitions also have to be general solutions. In the future work, we plan to investigate the patterns in these partition solutions and abstract these patterns to be used to classify binary inputs of any length $n$. 

An alternative approach to investigate general solutions for the quantized density classification problem is by applying the GA to search $S$ and $Z$ for different binary input sizes $n$. It is possible that such   solutions exist, but the GA may or may not be able to find them. This is because with the increase of the binary input length, the search space of $Z$ increases exponentially. However, it is also possible that the number of partition $Z$ solutions increases as the length of the binary inputs increases, hence the problem difficulty remains constant regardless of the binary inputs length. In either case, we can adjust GA parameters, such as $mR$ and $cR$, to improve the GA search efficiency. Another issue that we need to address using this approach is the hardware resources required to simulate the QPA evolution under large lattice sizes. These issues will be investigated in our future work.

Compared to the classical CA, which requires a lattice of length $L=n$ to solve the density classification problem for binary inputs of length $n$, the QPA requires lattice length $L=2^n$ under this quantized version of the density classification problem. However, the trade-off we have gained is a larger CA evolution space (complex Hilbert space) which allows many unitary operators to manipulate and solve the problem. In this research, our GA has found some of these solutions. Moreover, it is possible that the solutions are within some areas of the parameters space and can be simplified as 2-parameter unitary operators. We will test this hypothesis in our future work.

\section{Concluding Remarks}\label{conclusion}
Our investigation of using a quantum CA to solve the density classification problem not only produced some interesting results but also posed some challenging questions. First of all, we found that the QPA devised by Meyer can be used to solve a quantized version of the density classification problem for binary inputs of length 5 successfully. Secondly, we found that there are more than one quantum solution to this quantized problem and the GA system we designed can find many of them. Thirdly, we found that these quantum solutions propagate probability amplitudes in different speeds, depending mostly on a particular parameter $\theta$ of the unitary operator. Lastly, we found that there exists many boundary partitions in the 32 bases Hilbert space that separate the basis states classifying the density of the binary inputs as 0 from those classifying the density of the binary inputs as 1. When the partitions are paired with suitable unitary operators, they are able to solve the quantized density classification problem for binary inputs of length 5. 

However, scaling the approach to find general solutions to the quantized density classification problem requires more work. We have identified a couple of avenues: one studies the partition $Z$ patterns and the other improves GA scaling performance. We plan to investigate both approaches in our future work.  

Moreover, given the number of variables in the solutions ($\theta,\alpha,\beta,\gamma,Z,M$), it is not clear to us if we can reduce the number of parameters by increasing/decreasing other parameter values. For example, can we remove $Z$ by increasing $M$? Or can we remove $\alpha$ by increasing $M$? 
These are open questions that we plan to answer in our futuer work.

\section*{Appendix A}
\small
\texttt{\#include <stdio.h>\\
\#include <stdlib.h>\\
\#include <math.h>\\
\#include <complex.h>\\
double radian(double degree)\{return degree*M\_PI/180.0;\}\\
complex double e(double phi)\{return (cos(radian(phi))+ I * sin(radian(phi)));\}\\
double norm(complex double x)\{return creal(x)*creal(x)+cimag(x)*cimag(x);\}\\
/* Dimension of problem and boundaries for variables */\\
int D = 5;\\
/* Xl defines lower limit; Xu defines upper limit */\\
double Xl[5] = \{0.0, -180.0, -180.0, -180.0, 0.0\}; \\
double Xu[5] = \{90.0, 180.0, 180.0, 180.0, 4294967296.0\};\\
double SD[5] = \{0.05, 0.1, 0.1, 0.1, 20.0\};\\
/*majority binary code*/\\
const int majority[32]=\{0,0,0,0,0,0,0,1,0,0,0,1,0,1,1,1,0,0,0,1,0,1,1,1,0,1,1,1,1,1,1,1\};\\
/*majority gray code*/\\
//const int majority[32]=\{0,0,0,0,0,1,0,0,0,1,1,1,0,1,0,0,0,1,1,1,1,1,1,1,0,1,1,1,0,1,0,0\};\\
int func(double *rule) // rule=($\theta,\alpha,\beta,\gamma$,Z)\\
\{\\
\indent  // Correction of boundary constraint violations \\
\indent  for (i=0; i<D; i++)\\
\indent   \{\\
\indent\indent      if (rule[i] < Xl[i] || rule[i] > Xu[i]) \\
\indent\indent\indent rule[i] =  Xl[i] + (Xu[i] - Xl[i])*((double)rand()/((double)RAND\_MAX + 1.0));
\indent    \}\\ \\
\indent   // Converting Z to Zbinary\\       
\indent        int z\_binary[32]=\{0,0,0,0,0,0,0,0,0,0,0,0,0,0,0,0,0,0,0,0,0,0,0,0,0,0,0,0,0,0,0,0\};\\
\indent       int index;\\
\indent 	int L=32;\\
\indent      unsigned long Z = (unsigned long)rule[D-1];\\          
\indent         for(index=0; index < L; index++)\\
\indent    \indent     if(Z < 2)\\
\indent\indent            \{\\
\indent\indent\indent             z\_binary[index]=(int) Z;\\
\indent\indent\indent              index =L;\\
\indent\indent          \}\\
\indent \indent         else\\
\indent \indent             \{\\
\indent \indent \indent   z\_binary[index]=(int)(Z\%2);\\
\indent \indent \indent  Z=Z/2;\\
\indent \indent             \}         \\ 
\indent   // QPA simulation\\ 
\indent int i, j, step, start, k, n, o; \\ 
\indent int fitness=0;\\
\indent int M=2048;\\
\indent int SAMPLES = 32;\\
\indent double phiSum = 180.0;\\
\indent complex double U2[2][2];\\
\indent complex double ca[2049][32];\\
\indent  U2[0][0]=sin(radian(rule[0]))*e(rule[1]);\\
\indent U2[0][1]=cos(radian(rule[0]))*e(rule[2]);\\
\indent  U2[1][0]=cos(radian(rule[0]))*e(rule[3]);\\
\indent  U2[1][1]=sin(radian(rule[0]))*e(phiSum-rule[1]+rule[2]+rule[3]);\\
\indent  ca[0][0]=1.0;\\
\indent  for (j=1; j < L; j++)\\
\indent   \{\\
\indent  \indent    ca[0][j]=0.0;\\
\indent    \}\\
\indent  for (step=1; step <= M; step++)\\
\indent    \{\\
\indent \indent     start = step \%2;\\
\indent \indent     if(start == 1)\\
\indent \indent       \{ //odd step \\
\indent  \indent\indent        for(k = 0; k < L; k=k+2)\\
\indent  \indent\indent          \{\\
\indent  \indent\indent\indent            ca[step][k] = U2[0][0]*ca[step-1][k]+U2[0][1]*ca[step-1][k+1];\\
\indent  \indent\indent\indent            ca[step][k+1] = U2[1][0]*ca[step-1][k]+U2[1][1]*ca[step-1][k+1];\\
\indent \indent  \indent        \}\\
\indent   \indent     \}\\
\indent \indent     else /*shift right*/\\
\indent \indent       \{ \\
\indent \indent\indent        for(k = 1; k < L; k=k+2)\\
\indent   \indent    \indent      \{\\
\indent   \indent \indent    \indent         ca[step][k] = U2[0][0]*ca[step-1][k]+U2[0][1]*ca[step-1][k+1];\\
\indent  \indent \indent \indent             ca[step][k+1] = U2[1][0]*ca[step-1][k]+U2[1][1]*ca[step-1][k+1];\\
\indent   \indent     \indent      \}\\
\indent \indent  \indent       ca[step][L-1] = U2[0][0]*ca[step-1][L-1]+U2[0][1]*ca[step-1][0];\\
\indent  \indent  \indent      ca[step][0] = U2[1][0]*ca[step-1][L-1]+U2[1][1]*ca[step-1][0];\\
\indent  \indent      \}\\     \\
\indent \indent     /*check each of the 32 samples*/\\      
\indent  \indent    for(i = 0 ; i < SAMPLES; i++) \\
\indent   \indent \{\\
\indent \indent \indent double zero=0.0, one=0.0;\\ 
\indent  \indent \indent          int cell=(L-i)\%L; //i is the sample number\\          
\indent  \indent \indent         for(index=0; index < L; index++)\\
\indent \indent \indent             \{\\
\indent  \indent \indent \indent              if(z\_binary[cell]==0)\\
\indent \indent \indent \indent                 zero+=norm(ca[step][index]);\\
\indent \indent \indent \indent          cell=(cell+1)\%L;\\
\indent   \indent \indent          \}      \\   
\indent \indent \indent          one=1.0-zero;\\          
\indent \indent \indent          if(majority[i] == 0 \&\& zero > 0.5)\\
\indent \indent   \indent          \{\\
\indent \indent \indent   \indent            fitness++;\\
\indent \indent    \indent         \}\\
\indent  \indent\indent          else if(majority[i] == 1 \&\& one > 0.5)\\
\indent \indent \indent             \{\\
\indent \indent \indent \indent              fitness++;\\
\indent \indent \indent             \}\\
\indent \indent      \}//finish evaluating 32 samples\\ \\
\indent \indent     if( fitness == 32) \\
\indent \indent      \{       \\
\indent \indent\indent         for(n=0; n <= step; n++)\\
\indent \indent\indent     \indent        for(o=0; o < L; o++) \\
\indent   \indent\indent  \indent  \indent         printf(``\%f,'', norm(ca[n][o]));\\
\indent   \indent\indent      step = M;\\
\indent  \indent     \}\\
\indent    \} //close M steps loop\\
\indent  return fitness;\\
\} //close function
}

\end{document}